\documentclass[11pt,twoside]{article}
\usepackage{macroaaa-eng}
\usepackage{psfig}
\usepackage{graphicx}

\usepackage[T1]{fontenc} 

\usepackage{latexsym}
\usepackage{verbatim}

\usepackage{natbib,rotating,amssymb} 

\def\cmmoinsdeux{\mbox{ cm}^{-2}}

\def\microns{\mbox{ } \mu \mbox{m}}

\def\kpc{\mbox{ kpc}}

\def\kms{\mbox{ km\,s}^{-1}}

\def\Mdot{\frac{dM}{dt}}
\def\Msol{\mbox{ }M_{\odot}}
\def\Rsol{\mbox{ }R_{\odot}}
\def\Lsol{\mbox{ }L_{\odot}}

\def\Rstar{\mbox{ }R_{\star}}
\def\Tstar{\mbox{ }T_{\star}}

\def\Porb{P_{orb}}
\def\Pspin{P_{spin}}

\def\mags{\mbox{ magnitudes}}

\def\ergs{\mbox{ erg\,s}^{-1}}

\def\adeg{^{\circ}}

\def\amin{^\prime}

\def\nh{N_{\rm H}}

\def\ltsima{\; \buildrel < \over \sim \;}
\def\simlt{\lower.5ex\hbox{\ltsima}}            
\def\gtsima{\; \buildrel > \over \sim \;}
\def\simgt{\lower.5ex\hbox{\gtsima}}            

\begin{document}

\vskip 1.0cm
\markboth{S. Chaty}{{\it INTEGRAL} obscured sources and SFXTs}
\pagestyle{myheadings}

\vspace*{0.5cm}
\title{{\it INTEGRAL} sources: from obscured high mass X-ray binaries
to supergiant fast X-ray transients}

\author{Sylvain Chaty}
\affil{Laboratoire AIM, CEA/DSM - CNRS - Université Paris Diderot,
Irfu/Service d'Astrophysique, Centre de Saclay, Bât. 709,
F-91191 Gif-sur-Yvette Cedex, France, chaty@cea.fr}

\begin{abstract} A new type of high-energy binary system has been
  revealed by the {\it INTEGRAL} satellite.  These sources are being
  unveiled by means of multi-wavelength optical, near- and
  mid-infrared observations. Among these sources, two distinct classes
  are appearing: the first one is constituted of intrinsically
  obscured high-energy sources, of which IGR~J16318-4848 seems to be
  the most extreme example. The second one is populated by the
  so-called supergiant fast X-ray transients, with IGR~J17544-2619
  being the archetype. We first give here a general introduction on
  {\it INTEGRAL} sources, before reporting on multi-wavelength optical
  to mid-infrared observations of a sample constituted of 21 {\it
    INTEGRAL} sources. We show that in the case of the obscured
  sources our observations suggest the presence of absorbing material
  (dust and/or cold gas) enshrouding the whole binary system.  We
  finally discuss the nature of these two different types of sources,
  in the context of high energy binary systems, and give a scenario of
  unification of all these different types of high energy sources,
  based on their high energy properties.
\end{abstract}

\section{Prelude}

Last December it was the $45^{th}$ anniversary of the discovery of the
first X-ray extra-solar source --Sco X-1--, reported on December
$1^{st}$ 1962 by \citet{giacconi:1962}\footnote{R. Giacconi won the
  2002 Physics Nobel prize for this discovery.}. 40 years after these
early X-ray ages, the X-ray sky has been extensively observed, but as
we will see unexpected discoveries are still to expect!  Sco X-1, the
first X-ray source in the constellation of Scorpius, became the
prototype of Galactic high energy binary systems, in which it is now
commonly accepted that a compact objet (neutron star -NS- or black
hole -BH-) accretes from a so-called ``companion'', ``primary'' or
``secondary'' star.

We will begin by a short introduction on the distinction between low
mass and high mass X-ray binaries (LMXBs and HMXBs respectively).  We
will then continue with a review on {\it INTEGRAL} sources (called
IGRs in the following): the
general results obtained by {\it INTEGRAL}, the spatial distribution,
the modulation and absorption of sources discovered by this satellite.
We will then review some of the stellar winds properties, in order to
understand the basic theory on SFXTs, which will lead us to what we
call here the grand unification of sgXBs.

The following section will describe an extensive multi-wavelength
study based on a sample of IGRs, including the
classical SFXT IGR\,J17544-2619, the intermediate SFXT IGR\,J18483-0311,
and the obscured source IGR\,J16318-4848. We will finish with AX
J1749.1-2733, which might be the most obscured BeXRB, before
concluding and giving some perspectives.

\subsection{Galactic X-ray binaries}

There are now 300 X-ray binaries known in our Galaxy
(\citeauthor{liu:2006} \citeyear{liu:2006}; \citeauthor{liu:2007}
\citeyear{liu:2007}), that we can distinguish between LMXBs and HMXBs
according to the nature of the companion star hosted in the high enery
binary system.

\subsubsection{LMXBs:}

186 LMXBs (62\% of the total number of high energy binary systems): the
  companion has a spectral type later than B ($M<1\Msol$).  The mass
  transfer is done through Roche Lobe filling, via angular momentum
  loss through the accretion disc.  The LMXBs are concentrated in the
  Galactic bulge due to the fact that the companion stars, being less
  massive than the Sun, are old stars. The compact object can be
either a BH or a NS. When the NS is
magnetized, they are Z or Atoll sources according to their
hardness-intensity diagrammes.

\subsubsection{HMXBs:}

114 HMXBs (38\% of the total number of high energy binary systems): the
  companion has an OB spectral type ($M>10\Msol$).  The mass transfer
  occurs either via a decretion disk (Be systems), or a strong wind
  or even via Roche lobe filling (in
  the case of the supergiant X-ray binaries).  These systems are
  concentrated in the Galactic plane.
They are separated in 3 distinct groups according to the nature of their
companion star:

\begin{itemize}

\item The group of Be/X-ray Binaries (BeXBs, also called Be/X-ray
  transients) constitute the majority of HMXBs. The
  companion/donor star is in this case a main sequence Be spectral
  type star.  The compact object is a NS located on a wide and
  moderately eccentric orbit, and it is spending little time in close
  proximity to the dense circumstellar decretion disk surrounding the
  Be companion (\citeauthor{coe:2000} \citeyear{coe:2000};
  \citeauthor{negueruela:2004} \citeyear{negueruela:2004}).  Transient
  X-ray outbursts occur when the compact object passes through the
  Be-star disc, accreting from the low-velocity and high-density
  stellar wind.  It then exhibits hard X-ray spectra during the
  outburst.

\item The group of supergiant X-ray Binaries (sgXBs) is made of
  binaries hosting a donor supergiant early-type OB star.  The compact
  object is a NS orbiting deep inside the highly supersonic stellar
  wind \citep{kaper:2004}.  The sgXBs are still separated in two
  distinct groups, according to the accretion process:

\begin{itemize}

\item In the wind-fed systems, the X-ray luminosity is powered by
  accretion from the strong steady radiative stellar wind, creating a
  persistent X-ray source ($L_x\sim10^{35-36} \ergs$).  These systems
  exhibit large variations on short timescales (due to wind
  inhomogeneities), and a stable flux on the long run.  The compact
  object orbits on a close orbit ($P_{orb}<15d$) with low
  eccentricity.

\item In the Roche-lobe overflow systems (so-called classical ``bright'' sources), the matter flows via the inner Lagrangian point forming an accretion disc,
giving a high X-ray luminosity ($L_x\sim10^{38} \ergs$) during outbursts.
We point out that Cyg X-1 is the only sgXB hosting a confirmed BH.

\end{itemize}

\item Finally, the last group of HMXBs is constituted of all the other
  main sequence or giant type companion HMXBs (including the 
symbiotic systems).

\end{itemize}

\subsection{The Corbet Diagramme}

In 1986, \cite{corbet:1986} reported a plot of the NS spin period
versus the orbital period of HMXBs. This plot, now known as the "Corbet
Diagramme", shows that the 3 types of HMXB pulsators (BeXBs, wind-fed
and Roche-lobe filling sgXBs) segregate into different regions
of this diagramme, owing to the complex feedback processes 
between modulation periods and dominant accretion mechanism.

A correlation is observed in Be systems, due to the fact that they 
accrete significant angular momentum to be able to form an accretion disc.
On the other hand, 
no correlation is observed in supergiant systems, because of the 
low net angular momentum of accreted matter: if an accretion disc
is formed during accretion of matter, it must be of transitory nature.

\section{And then {\it INTEGRAL} arrived...}

The {\it INTEGRAL} observatory is an ESA satellite launched on 17 October
2002 by a PROTON rocket on an excentric orbit. It is hosting 4
instruments: 2 $\gamma$-ray coded-mask telescopes --the imager IBIS
and the spectro-imager SPI, observing in the range 10 keV-10 MeV, with
a resolution of $12\amin$ and a field-of-view of $19\adeg$-- a
coded-mask telescope JEM-X (3-100 keV), and an optical telescope
(OMC).

One of the important results of {\it INTEGRAL} has been
obtained during the deep observation of
our Galaxy the Milky Way, allowing to show that the diffuse X-ray
background, below 80 keV, could be entirely resolved into X-ray point
sources \citep{lebrun:2004}.

\subsection{The $\gamma$-ray sky seen by {\it INTEGRAL}}

The $\gamma$-ray sky seen by {\it INTEGRAL} is very rich, since 499 sources
have been detected by {\it INTEGRAL}, reported in the $3^{rd}$ IBIS/ISGRI soft
$\gamma$-ray catalogue, spanning 3.5 years of observations in the 20-100
keV domain \citep{bird:2007}.  Among them, 214 were discovered by
{\it INTEGRAL}, while the remaining 285 were previously known.

Among these sources, there are 147 XRBs (29\% of the total number of {\it INTEGRAL}
sources), 163 AGNs (33\%), 27 CVs (5\%), and 20 sources of other type
(4\%): 12 SNRs, 2 globular clusters, 2 SGRs and 1 GRB. 129 objects
still remain unidentified (26\%).  The XRBs are separated in 82
LMXBs (16\%) and 78 HMXBs (16\%).  Among the HMXBs, there are 24 BeXBs
(31\% of the total of HMXBs) and 19 sgXBs (24\% of the total of HMXBs).

It is interesting to follow the evolution of the ratio between BeXBs
and sgXBs.  During the pre-{\it INTEGRAL} era, HMXBs were mostly BeXBs
systems.  For instance, in the catalogue of 130 HMXBs by
\cite{liu:2000}, there were 54 BeXBs (42\% of the total number of
HMXBs) and 7 sgXBs (5\%).  Then, the situation changed with the first
HMXBs identified by {\it INTEGRAL}: in the catalogue of HMXBs of
\cite{liu:2006}, out of 114 HMXBs (+128 in Magellanic Clouds), there
were 60\% of BeXBs and 32\% of sgXBs firmly identified.  Therefore,
the ratio of BeXBs/HMXBs increased by a small factor of 1.5, while the
one of sgXBs/HMXBs increased by a much higher factor of 6.

We will see later that the two highlights of {\it INTEGRAL} in the XRB
domain are first the 
emergence of an obscured population of sgXBs, and then the
emergence of the SFXT class (12 candidates).

In the 3 following subsections we follow the study reported in the
 exhaustive and excellent paper of \cite{bodaghee:2007}.

\subsection{The Galactic spatial distribution}

We now examine the impact of stellar evolution of massive
binaries on the formation of binary systems, by looking at the spatial
distribution of binary systems with known distances on a Galactic
spiral 4-arm model (based on locations of star-forming regions --SFRs--
and related complexes: OB stars, molecular clouds, HII regions,
diffuse ionised gas; see \citeauthor{russeil:2003}
\citeyear{russeil:2003}).

The LMXBs, hosting old companion stars which had the time to migrate
off the Galactic plane (|b| > $3-5\adeg$), are found to be
concentrated in or near the Galactic bulge where old globular clusters
reside, peaking at the center and decreasing gradually with the
galactocentric radius, suggesting an association with the Galactic
bar.

On the contrary, HMXBs, hosting young companion stars, are encountered
in recent stellar formation sites, in the outer disk and arms where
young stars are formed, following the HII/CO distribution.
Underabundant in central few kpc, their uneven distribution along the
Galactic plane reflects the Galactic spiral structure. Indeed, they
coincide with the active young massive SFRs, peaking
at $l \sim \pm 30\adeg$ towards tangential directions of inner spiral
arm tangents (Norma and Scutum/Sagittarius) and towards a molecular
ring located at $\sim 3$~kpc from the Galactic center.

The spread of latitude distributions from the Galactic plane is larger
in LMXBs than in HMXBs due to the relative young HMXB companions.
Such evolutionary signatures had been already noticed by Ginga
\citep{koyama:1990} and RXTE \citep{grimm:2002} but on smaller
samples. However we point out that the exposure map is heterogeneous.
According to their angular distribution and transience, the population
of yet unclassified sources is likely composed primarily of Galactic
LMXBs/CVs. The distribution of miscellaneous sources is similar to
HMXBs.

Since the propagation of density waves promotes stellar formation in
spiral arms \citep{lin:1969}, the distribution of HMXBs is offset with
directions of spiral arm tangents because it requires $\sim 10$~Myr
before one of the stars in a binary system collapses into a NS/BH.
The Galactic rotation induces changes in the apparent
positions of the arms, causing a delay between star formation epoch
and time of maximum number of HMXBs.  Therefore the currently active
star-forming sites should be about $\sim 40 \adeg$ away from regions which
were active 10 Myr ago and produced the current HMXBs
(\citeauthor{lutovinov:2005a} \citeyear{lutovinov:2005a}; 
\citeauthor{dean:2005} \citeyear{dean:2005}; 
\citeauthor{bodaghee:2007} \citeyear{bodaghee:2007}).

\subsection{Absorption}

IGRs column densities are higher (by a factor of $\sim 4$) than
expected from the radio maps. Previously known sources had
$<\nh>=1.2\times10^{22}\cmmoinsdeux$, while IGRs exhibit
$<\nh>=4.8\times10^{22}\cmmoinsdeux$.  The Galactic IGRs are HMXBs with
high $\nh$ ($\sim 10^{23}\cmmoinsdeux$).  The question which
arises is therefore:
are the Galactic IGRs intrinsically absorbed due to the geometry of
the  absorbing material, or extrinsically due to their location
along the dusty Galactic plane?  While the line-of-sight absorption
shows that there are potential clustering or asymmetries in the local
distribution of matter, the highest value of Galactic $\nh$ is $\sim
3\times10^{22}\cmmoinsdeux$ \citep{dickey:1990}.  It seems therefore
that the high absorption is intrinsic to IGRs.  ISGRI (>20 keV)
is immune to absorption that prevented discovery of absorbed sources
with earlier soft X-ray telescopes.

The most heavily-absorbed Galactic sources ($\nh>10^{23}\cmmoinsdeux$)
are localized in the Norma Arm region, the most active formation site
of young supergiant stars \citep{bronfman:1996} --precursors to the
absorbed HMXBs-- and the Galactic Bulge and Scutum/Sagittarius arms.

\subsection{Modulations}

Strong magnetic fields in NS XRBs produce non-spherically symmetric
emission patterns: due to misaligned magnetic and rotation axes,
pulsations are seen in the X-ray light curve.  Most IGRs have spin
periods $\Pspin$ = 100 - 1000s, i.e. 10 times longer than <$\Pspin$> of pre-IGRs.

IGRs exhibiting extreme modulations are:

\begin{itemize}

\item IGR\,J00291+5934, with $\Pspin$ = 1.7 ms, is the 
fastest accretion-powered
  pulsar ever observed \citep{galloway:2005},

\item IGR\,J16358 - 4726, with $\Pspin$ = 6000 s, is the slowest
  NS rotator 
(\citeauthor{lutovinov:2005a} \citeyear{lutovinov:2005a}; 
\citeauthor{patel:2007} \citeyear{patel:2007}).

\end{itemize}

Why do HMXB IGRs have longer pulse periods?  sgXBs are wind-fed
systems with strong magnetic fields that tend to have the longest
pulse periods (e.g. \citeauthor{corbet:1984} \citeyear{corbet:1984}).
{\it INTEGRAL} and {\it XMM-Newton} feature long orbital periods around Earth.
The distribution of IGR $\Porb$ exhibit a bimodal shape similar to
the one known before {\it INTEGRAL}, with 2 populations: LMXBs (and CVs,
SNRs...)  exhibit shorter $\Porb$, and HMXBs longer $\Porb$.  The
Corbet $\Pspin$ - $\Porb$ diagramme shows that the majority of IGRs is
located among other known wind-fed sgXBs.  BeXBs have 
in general longer $\Porb$
than sgXBs (this was already known, see e.g.
\citeauthor{stella:1986} \citeyear{stella:1986}).  {\it INTEGRAL} is
therefore not just finding new HMXBs pulsars, but predominantly
long-period sgXBs.

\subsection{Modulation vs Absorption}

The accretion affects the $\Pspin$ of a NS, depending on the velocity at
the corotation radius ($V_c$) where the magnetic field regulates the motion
of matter:

\begin{itemize}

\item if $V_c$ > $V_{Kepler}$: the material is spun away, taking angular
  momentum, and the NS slows down due to "propellor mechanism"
  \citep{illarionov:1975}.

\item if $V_c$ < $V_{Kepler}$: the material is accreted onto the NS,
  which either spins up or down depending on the direction of the
  angular momentum with respect to the NS spin \citep{waters:1989}.

\end{itemize}

Since the spherically-symmetric accretion on a NS is driven
by the wind of the companion star, the HMXB pulsar spin rate is
regulated by the stellar wind angular momentum of the companion star.

The wind density of a supergiant star has the form $\rho(r) \propto
r^{-2}$.  For Be stars, because stellar winds have dense and slow
equatorial outflows and thin fast polar winds \citep{lamers:1987}, the
wind density drops faster: $\rho(r) \propto r^{-3}$
\citep{waters:1988}. In addition, there are stronger density and
velocity gradients inside the capture radius of the NS, in both radial
and azimutal directions.  Wind-fed accretion is therefore more
efficient at delivering angular momentum to the NS in BeXBs than it
is in sgXBs \citep{waters:1989}.

Given the density structures, and assuming a steady accretion rate
with angular momentum of same direction as the NS spin, it reaches an
equilibrium value $P_{eq}$ propto $\rho^{-3/7}$. However, the current
$\Pspin$ of NS in sgXBs are much longer than predicted and are
closer to $P_{eqs}$ of the stellar winds while the star was still on
the main sequence \citep{waters:1989}. On the other hand, the
equilibrium $\Pspin$ in Be systems is constantly adjusting to the
changing conditions in the winds. As with sg systems, pulsars in Be
systems are not currently spinning at $P_{eq}$ but reflect the values
of an earlier evolutionary stage \citep{king:1991}.
The transport of positive angular momentum through the wind is so
inefficient that it can not spin up the pulsar to its expected
equilibrium $P_{eqs}$. However the pulsar can be spun down by the
"propellor mechanism".

\subsubsection{A new tool:}

Galactic HMXB IGRs segregate into distinct regions of $\Pspin$ -
$\nh$ diagramme: sgXBs have higher average $\nh$ and longer average
$\Pspin$ compared to BeXBs. There are only 2 sgXBs with $\Pspin$ <
50s: Cen X-3 a Roche-lobe overflow system and OAO 1657 - 415, which
seem to be intermediate systems, since they show transitions from
wind-fed to disk-fed system.


In view of all what has been described until now, looking at the
location of IGRs in a $\Pspin$-$\nh$ diagramme constitutes a new tool to
distinguish between sg and BeXBs when only $\nh$ and either $\Pspin$
or $\Porb$ are known.

\begin{itemize}

\item IGR\,J19140+0951 ($P_{orb} \sim 13d$, $\nh \sim 10^{23}
  \cmmoinsdeux$): positioned among sgXBs, the sgOB spectral type has
  been confirmed by IR observations (see e.g. \citeauthor{chaty:2008}
  \citeyear{chaty:2008}; and see also \citeauthor{prat:2008}
  \citeyear{prat:2008} for the dependence of $\nh$ along the phase);

\item IGR\,J16358 - 4726: exhibiting an unusually long $\Pspin$ 
suggesting a magnetar nature \citep{patel:2007}, and located
in the sgXB part, the spectral type has been confirmed by \cite{chaty:2008}.


\end{itemize}

\section{Now that {\it INTEGRAL} is orbiting... let the show go on!}

The {\it INTEGRAL} observatory has performed a detailed survey of the
Galactic plane.  The ISGRI detector on the IBIS imager has discovered
many new high energy celestial objects, most of which have been
reported in \cite{bird:2007}\footnote{See an updated list at {\em
    http://isdc.unige.ch/$\sim$rodrigue/html/igrsources\-.html}}.  The
most important result of {\it INTEGRAL} to date is the discovery of
many new high energy sources -- concentrated in the Galactic plane,
and towards the Norma arm, a region of our Galaxy full of star forming
regions, -- exhibiting common characteristics which previously had
rarely been seen (see e.g. \citeauthor{chaty:2005a}
\citeyear{chaty:2005a}). Many of them are high mass X-ray binaries
(HMXBs) hosting a NS orbiting around an OB companion, in
most cases a supergiant star. They divide into two classes: some of
the new sources are very obscured, exhibiting a huge intrinsic and
local extinction, --the most extreme example is the highly absorbed
source IGR~J16318-4848 \citep{filliatre:2004}--, and the others are
HMXBs hosting a supergiant star and exhibiting fast and transient
outbursts -- an unusual characteristic among HMXBs.  These are
therefore called Supergiant Fast X-ray Transients (SFXTs,
\citeauthor{negueruela:2006a} \citeyear{negueruela:2006a}),
with IGR~J17544-2619
being their archetype \citep{pellizza:2006}.

Nearly all the {\it INTEGRAL} HMXBs for which both spin and orbital
periods have been measured are located in the upper part of the Corbet
diagramme \citep{corbet:1986}.  They are wind accretors, typical of
supergiant HMXBs, and X-ray pulsars exhibiting longer pulsation
periods and higher absorption (by a factor $\sim4$) as compared to the
average of previously known HMXBs \citep{bodaghee:2007}. This extra
absorption might be due to the presence of a cocoon of dust/cold gas 
enshrouding the whole binary system in the case of the
obscured sources.  The intrinsic properties of the supergiant companion star 
could therefore explain some properties of these
sources.  However, differences exist between obscured
sources and SFXTs, which might be explained by the geometry of the
binary systems, and/or the extension of the wind/cocoon enshrouding
either the companion star or the whole system.  Indeed, obscured
sources are naturally explained by a compact object orbiting inside a
cocoon of dust and/or cold gas, while the fast X-ray behaviour of
SFXTs needs a clumpy stellar wind environment, to account for fast and
transient accretion phenomena (see Figure \ref{figure:obscured-sfxt},
left and right panels respectively, and \citeauthor{chaty:2006c}
\citeyear{chaty:2006c}).

How to reveal the nature of these new newly discovered IGRs?
High-energy observations are not sufficient to reveal their nature,
since the {\it INTEGRAL}/ISGRI localisation ($\sim 2\amin$) is not
accurate enough to unambiguously pinpoint and identify the counterpart
at other wavelengths. Once X-ray satellites such as {\it XMM-Newton},
{\it Chandra}, or {\it Swift} provide an arcsecond position, the hunt
for the optical counterpart of the source is open.  However, the high
level of absorption due to interstellar dust and gas towards the
Galactic plane/centre makes the near-infrared (NIR) domain more
efficient for identifying these sources.

On the other hand, multi-wavelength observations allow us to study
various components of the system, emitting in various wavelength
domains, depending on the nature of the sources.  There is a
complementarity of telescopes from space to ground, starting with the
discovery of the high energy source by X/$\gamma$-ray satellites such as {\it
  INTEGRAL}, and then its localization in X-rays by {\it XMM-Newton}, {\it
  Swift}, or even {\it Chandra}, bringing a sub-arcsecond
localisation.  After this localization, the counterpart can be further
looked for in radio (VLA), optical and/or infrared wavelengths (for
instance using ESO facilities).

In this section we now report on multi-wavelength observations of a
sample of IGRs, belonging to both classes described above, and
we then give general results on IGRs, before
discussing them and concluding.

\subsection{Multi-wavelength observations of  {\it INTEGRAL} Sources} \label{observations-IGRs}

To better characterize this population, \cite{chaty:2008},
\cite{rahoui:2008} and \cite{tomsick:2007} studied a sample of 21 IGRs
belonging to both classes described above.  IGRs of
this sample are X-ray pulsars, with high $\Pspin$ from 139 to 5880s
and $\Porb$ ranging from 4 to 14 days.  They are therefore HMXBs wind
accreting supergiants, according to the Corbet diagramme.  The
multiwavelength observations were performed from 2004 to 2008 at the
European Southern Observatory (ESO), using Target of Opportunity (ToO)
and Visitor modes, in 3 domains: optical ($400-800$\,nm) with EMMI,
NIR ($1-2.5 \microns$) with SOFI, both instruments at the focus of the
3.5m New Technology Telescope (NTT) at La Silla, and mid-infrared
(MIR, $5-20 \microns$) with the VISIR instrument on Melipal, the 8m
Unit Telescope 3 (UT3) of the Very Large Telescope (VLT) at Paranal
(Chile). They also used data from the GLIMPSE survey of {\it Spitzer}.  With
these observations they performed accurate astrometry, identification,
photometry and spectroscopy on this sample of IGRs,
aiming at identifying their counterparts and the nature of the
companion star, deriving their distance, and finally characterising
the presence and temperature of their circumstellar medium, by fitting
their spectral energy distribution (SED).  

Some results are reported in Table \ref{table:results}.  Before
describing some of these multi-wavelength results in detail, we first
mention the main results of this study. 15 of these IGR sources were
identified as HMXBs, and among them 12 HMXBs containing massive and
luminous early-type companion stars. By combining MIR photometry, and
fitting their optical--MIR SEDs,
\cite{rahoui:2008} showed that (i) most of these sources exhibit an
intrinsic absorption and (ii) three of them exhibit a MIR excess,
which they suggest to be due to the presence of a cocoon of dust and/or
cold gas enshrouding the whole binary system, with a temperature of
$T_d \sim 1000K$, extending on a radius of $R_d \sim 10\Rstar$ (see also
\citeauthor{chaty:2006c} \citeyear{chaty:2006c}).

\begin{sidewaystable*}
  \begin{center}
    \caption{Results on the sample of IGRs; more
details are given in \cite{chaty:2008}. We indicate respectively the
name of the sources, the region of the Galaxy in the direction 
which they are located (No: Norma, GC: Galactic Centre), their spin and orbital period,
(in seconds and days respectively), and absorption derived from observations of
the interstellar, optical-IR, and X-ray derived column density
respectively (in units of $10^{22}\cmmoinsdeux$), their spectral type, their
nature and reference on these sources.
Type abbreviations: 
AGN = Active Galactic Nucleus,
B = Burster,
BHC = Black Hole Candidate,
CV = Cataclysmic Variable,
D = Dipping source,
H = High Mass X-ray Binary system, 
IP = Intermediate polar, 
L = Low Mass X-ray Binary,
O = Obscured source,
P = Persistent source,
S = Supergiant Fast X-ray Transient,
T: Transient source,
XP: X-ray Pulsar.
Reference are:
c: \cite{chaty:2008},
co: \cite{combi:2006},
f: \cite{filliatre:2004},
h: \cite{hannikainen:2007},
m1: \cite{masetti:2004}
m2: \cite{masetti:2006}
n1: \cite{negueruela:2005},
n2: \cite{negueruela:2006a},
p: \cite{pellizza:2006},
t: \cite{tomsick:2006a}.
}
\label{table:results}
\vspace{1em}
    \renewcommand{\arraystretch}{1.2}
    \begin{tabular}{cccccccccc} 
\hline
Source & Reg & P$_s$(s) & P$_o$(d) & $\nh{is}$ & $\nh{IR}$ & $\nh{X}$ & SpT & Type & Ref \\
\hline
IGR\,J16167-4957 & No & & & 2.2 & 0.23 & 0.5 & A0 & CV/IP & t,m2 \\
\hline 
IGR\,J16195-4945 & No & & & 2.18 & 2.9 & 7 & OB & H?/S?/O & t \\
\hline
IGR\,J16207-5129 & No & & & 1.73 & 2.0 & 3.7 & BOI & H/O & t,m2 \\
\hline
IGR\,J16318-4848 & No & & & 2.06 & 3.3 & 200 & sgB[e] & H/O/P & f \\
\hline
IGR\,J16320-4751 & No & 1250 & 8.96(1) & 2.14 & 6.6 & 21 & sgOB & H/XP/T/O & c \\
\hline
IGR\,J16358-4726 & No & 5880 & & 2.20 & 3.3 & 33 & sgB[e]? & H/XP/T/O & c \\
\hline
IGR\,J16393-4643 & No & 912 & 3.6875(6) & 2.19 & 2.19 & 24.98 & BIV-V? & H/XP/T & c \\
\hline 
IGR\,J16418-4532 & No & 1246 & 3.753(4) & 1.88 & 2.7 & 10 & sgOB? & H/XP/S & c \\
\hline
IGR\,J16465-4507& No & 228 & & 2.12 & 1.1 & 60 & B0.5I & H/S & n1 \\
\hline
IGR\,J16479-4514 & No & & & 2.14 & 3.4 & 7.7 & sgOB & H/S? & c \\
\hline
IGR\,J16558-5203 & - & - & - & - & - & - & Sey1.2 & AGN & m2 \\
\hline 
IGR\,J17091-3624 & GC & & & 0.77 & 1.03 & 1.0 & L & L/BHC & c \\
\hline 
IGR\,J17195-4100 & GC & & & 0.77 & & 0.08 & & CV/IP & t,m2 \\
\hline 
IGR\,J17252-3616 & GC & 413 & 9.74(4) & 1.56 & 3.8 & 15 & sgOB & H/XP/O & c \\
\hline
IGR\,J17391-3021 & GC & & & 1.37 & 1.7 & 29.98 & O8Iab(f) & H/S/O & n2 \\
\hline
IGR\,J17544-2619 & GC & & $\geq$70? & 1.44 & 1.1 & 1.4 & O9Ib & S & p \\ 
\hline
IGR\,J17597-2201 & GC & & & 1.17 & 2.84 & 4.50 & L & L/B/D/P & c \\
\hline
IGR\,J18027-1455 & - & - & - & - & - & - & Sey1 & AGN & m1,co \\
\hline 
IGR\,J18027-2016 & GC & 139 & 4.5696(9) & 1.04 & 1.53 & 9.05 & sgOB & H/XP/T & c \\
\hline
IGR\,J18483-0311 & GC & 21.05 & 18.55 & 1.62 & 2.45 & 27.69 & sgOB? & H/XP & c \\
\hline 
IGR\,J19140+0951 &    & & 13.558(4) & 1.68 & 2.9 & 6 & sgB0.5I & H/O & h \\
\hline
\end{tabular}
  \end{center}
\end{sidewaystable*}

\subsection{Supergiant Fast X-ray Transients}

\subsubsection{General characteristics:}

Supergiant Fast X-ray transients (also called ``SFXTs'') are high-mass
X-ray binary systems hosting NS orbiting around sgOB companions.  They
constitute a new class of sources identified among the recently
discovered IGRs, exhibiting the following common
characteristics: they exhibit rapid outbursts lasting only for hours,
a faint quiescent emission, high energy spectra requiring a BH or NS
accretor, and supergiant OB companion stars.  The flares have the
following characteristics: their rise last for tens of minutes, the
total duration is of $\sim$ 1 hour, their frequency of $\sim$ 7 days,
and their luminosity $L_x \sim 10^{36} \ergs$ at the outburst peak.
Furthermore, they exhibit a faint quiescent emission.  These sources
therefore exhibit general properties different from ``classical''
HMXBs.

   \subsubsection{IGR~J17544-2619 --Archetype of the SFXTs--:}

\begin{figure}
  \includegraphics[height=.37\textheight,angle=-90]{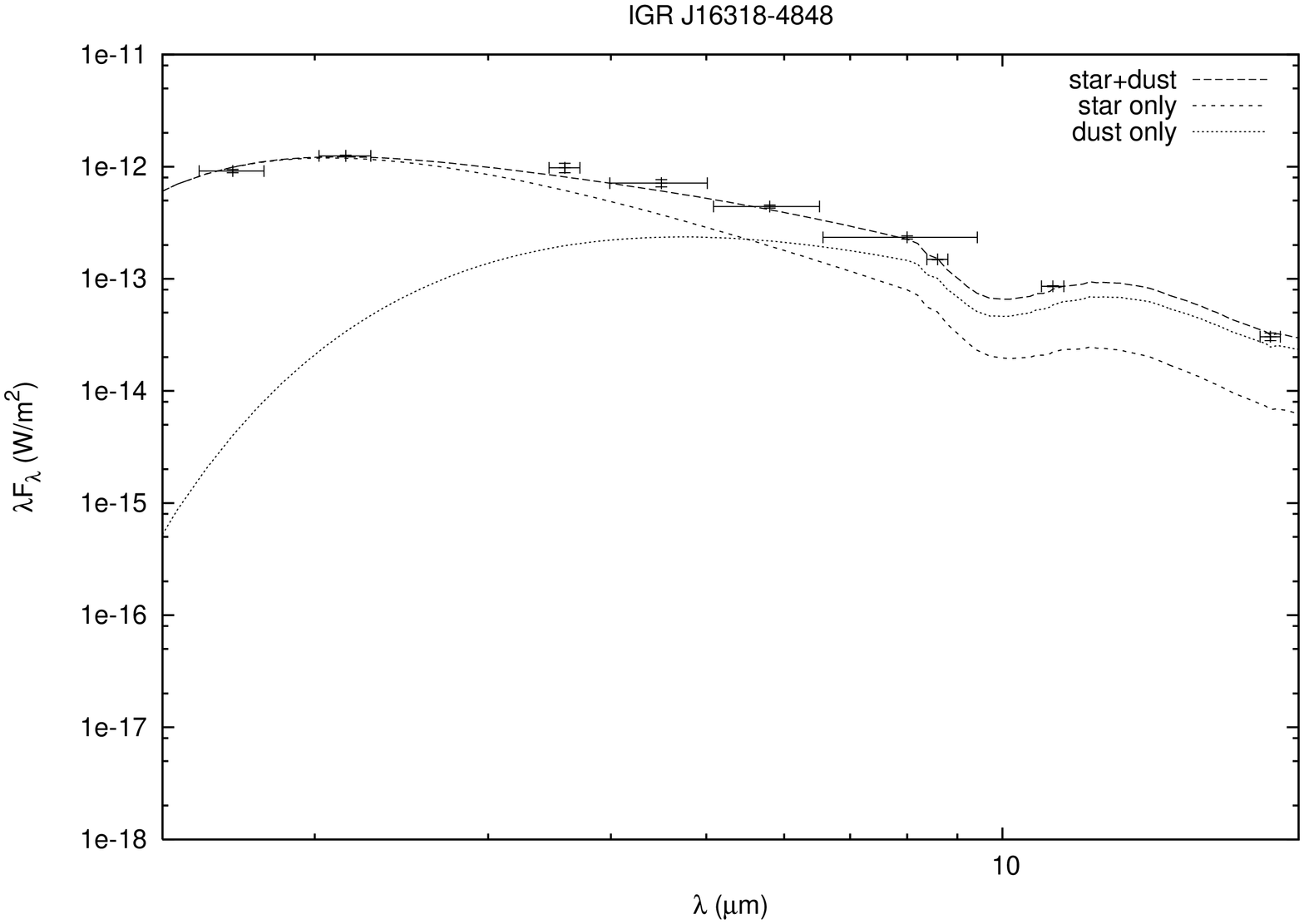}
  \includegraphics[height=.37\textheight,angle=-90]{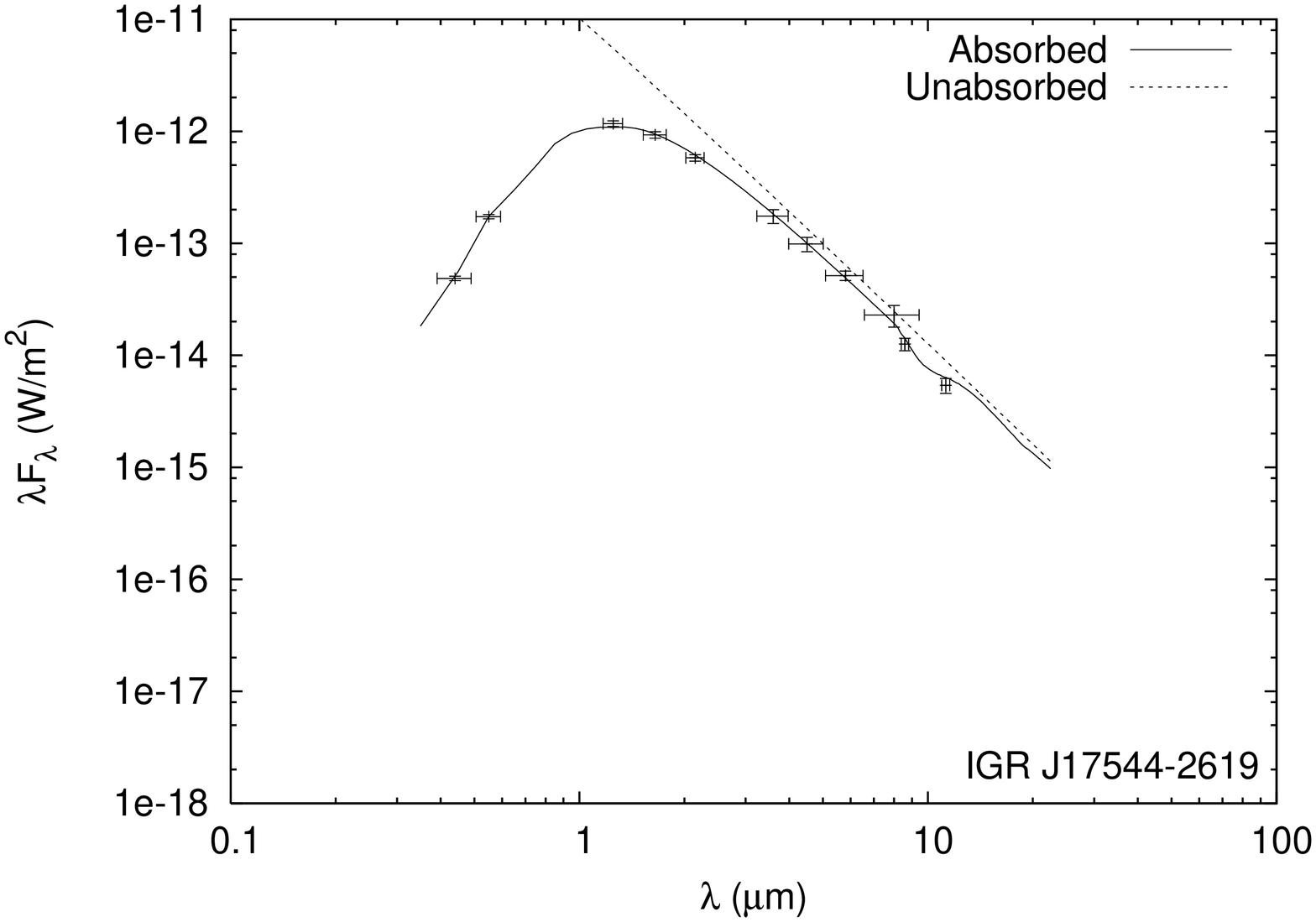}
  \caption{\label{figure:igrj16318-igrj17544} Optical to MIR SEDs of
  IGR~J16318-4848 (left) and IGR~J17544-2619 (right), including data
  from ESO/NTT, VISIR on VLT/UT3 and {\it Spitzer} \citep{rahoui:2008}.
  IGR~J16318-4848 exhibits a MIR excess, interpreted by
  \cite{rahoui:2008} as the signature of a strong stellar outflow
  coming from the sgB[e] companion star \citep{filliatre:2004}.  On the
  other hand, IGR~J17544-2619 is well fitted with only a stellar
  component corresponding to the O9Ib companion star spectral type
  \citep{pellizza:2006}.}
\end{figure}

IGR~J17544-2619, a bright recurrent transient X-ray source
   discovered by {\it INTEGRAL} on 17 September 2003
   \citep{sunyaev:2003b}, seems to be their archetype. Observations
   with {\it XMM-Newton} have shown that it exhibits a very hard X-ray
   spectrum, and a relatively low intrinsic absorption ($\nh \sim 2
   \times 10^{22}\cmmoinsdeux$, \citeauthor{gonzalez-riestra:2004}
   \citeyear{gonzalez-riestra:2004}).  Its bursts last for hours, and
   inbetween bursts it exhibits long quiescent periods, which can
   reach more than 70 days. The X-ray behaviour is complex on long,
   mean and short-term timescales: rapid flares are detected by
   {\it INTEGRAL} on all these timescales, on pointed and 200s binned
   lightcurve \citep{zurita-heras:2008a}. The compact object is
   probably a NS \citep{intzand:2005}.  \cite{pellizza:2006}
   managed to get optical/NIR ToO observations only one day after the
   discovery of this source. They identified a likely counterpart
   inside the {\it XMM-Newton} error circle, confirmed by an accurate
   localization from {\it Chandra}.  Spectroscopy showed that the
   companion star was a blue supergiant of spectral type O9Ib, with a
   mass of $25-28 \Msol$, a temperature of $T\sim 31000$~K, and a
   stellar wind velocity of $265 \pm 20 \kms$ (which is faint for O
   stars): the system is therefore an HMXB \citep{pellizza:2006}.
   \cite{rahoui:2008} combined optical, NIR and MIR observations and
   showed that they could accurately fit the observations with a model
   of an O9Ib star, with a temperature $\Tstar \sim 31000$~K and a radius
   $\Rstar = 21.9 \Rsol$. They derived an absorption A$_v = 6.1
   \mags$ and a distance D~$=3.6$~kpc. Therefore the source does not exhibit
   any MIR excess, it is well fitted by a unique stellar component
   (see Figure \ref{figure:igrj16318-igrj17544}, right panel,
   \citeauthor{rahoui:2008} \citeyear{rahoui:2008}).

   In summary, IGR~J17544-2619 is an HMXB at a distance of
   $\sim$3.6~kpc, constituted of an O9Ib supergiant, with a mild
   stellar wind and a compact object which is probably a NS,
   without any MIR excess.

\subsubsection{Classification of SFXTs:}

We can divide the SFXTs in two groups, according to the duration and
frequency of their outbursts, and their $\frac{L_{max}}{L_{min}}$ ratio.
The classical SFXTs exhibit a very low quiescence $L_X$, and a high
variability, while intermediate SFXTs exhibit a higher <$L_X$>, a
lower $\frac{L_{max}}{L_{min}}$ and a smaller variability, with longer
flares.  SFXTs might appear like persistent sgXBs with <$L_X$> below the
canonical value of $\sim 10^{36} \ergs$, and flares superimposed.  But
there might be some observational biases, therefore the distinction
between SFXTs and sgXBs is not well defined yet.
While the typical hard X-ray variability factor (the ratio between the
deep quiescence to outburst flux) is less than 20 in
classical/absorbed systems, it is higher than 100 in SFXTs (some
sources can exhibit flares in a few minutes, like for instance
XTE\,J1739-302 \& IGR\,J17544-2619).  The
intermediate SFXTs exhibit smaller variability factors. \\

\underline{SFXT behaviour: clumpy wind accretion?}

Such sharp rises exhibited by SFXTs are incompatible with the orbital
motion of a compact object through a smooth medium
(\citeauthor{negueruela:2006a} \citeyear{negueruela:2006a},
\citeauthor{smith:2006} \citeyear{smith:2006},
\citeauthor{sguera:2005} \citeyear{sguera:2005}).  Instead, flares
must be created by the interaction of the accreting compact object
with the dense clumpy stellar wind (representing a large fraction of
stellar $\Mdot$).  In this
case, the flare frequency depends on the system geometry, and the
quiescent emission is due to accretion onto the compact object of
diluted inter-clumps medium, explaining the very low quiescence level
in classical SFXTs.

To explain the emission of sgXBs/SFXTs, 
\cite{negueruela:2008} and \cite{walter:2007} invoke the
existence of two zones around the supergiant star, of high and low
clump density respectively.  This would naturally explain the smooth
transition between sgXBs and SFXTs, and the existence of intermediate
systems; the main difference between the classical sgXBs and the SFXTs being
in this scenario the NS orbital radius:

\begin{itemize}

\item The obscured sgXBs (persistent and luminous systems) would have 
short and circular orbit inside the zone of high clump density ($R_{orb} \sim 2\Rstar$).

\item The intermediate SFXT would have short orbits, circular or eccentric, 
and possible periodic outbursts.

\item The classical SFXT would have larger and eccentric orbital radius.

\end{itemize}

\underline{Macro-clumping scenario}

Each SFXT outburst is due to the accretion of a single clump, assuming that 
the X-ray lightcurve is a direct tracer of the wind density distribution.
The typical parameters in this scenario are:

\begin{itemize}

\item a compact object with large orbital radius: $10 \Rstar$,

\item a clump size of a few tenths of $\Rstar$,

\item a clump mass of $10^{22-23}g$ (for $\nh=10^{22-23}\cmmoinsdeux$),

\item a mass loss rate of $10^{-(5-6)} \Msol/yr$,

\item a clump separation of order $R_{\star}$ at the orbital radius,

\item a volume filling factor: 0.02->0.1

\end{itemize}

The flare to quiescent count rate ratio is directly related to the $\frac{clumps}{inter-clump}$ density ratio, which ranges between:

\begin{itemize}

\item 15-50 for intermediate SFXTs, and

\item $10^{2-4}$ for "classical" SFXTs.

\end{itemize}

A very high degree of porosity (macroclumping) is required to reproduce
the observed outburst frequency in SFXTs, in good agreement with UV line
profiles and line-driven instabilities at large radii 
(\citeauthor{oskinova:2007} \citeyear{oskinova:2007};
\citeauthor{runacres:2005} \citeyear{runacres:2005}; 
\citeauthor{walter:2007} \citeyear{walter:2007}). 
The number of clumps vs radius in a ring of width $2r_{acc}$ and 
height $2_{racc}$ is given in \cite{oskinova:2007}, for a
velocity law with beta=0.8 and porosity parameter $L_0=0.35$. \\

\underline{Difference sgXB/SFXT}

A basic model of porous wind predicts a substantial change in the
properties of the wind "seen by the NS" at a distance $r \sim 2
\Rstar$ (vertical asymptote in Figure 1 of
\citeauthor{negueruela:2008} \citeyear{negueruela:2008}), where we
stop seeing persistent X-ray sources. There are 2-regimes:

\begin{itemize}

\item either the NS sees a large number of clumps, because it is
  embedded in a quasi-continuous wind;

\item or the number density of clumps is so small that the NS is effectively 
orbiting in an empty space.

\end{itemize}

sgXBs can only lie within the two vertical dot-dashed lines \citep{negueruela:2007}. \\

\underline{Classes of wind accretors}

In \cite{negueruela:2007} Figure 2 represents the stellar wind of high
(coloured area) and low (blank area) clump density respectively. The
coloured are represents the left part of the asymptote in Figure 1 of
the same paper. The HMXB configurations are:

\begin{itemize}

\item SFXTs on the left (circular orbit, NS outside the high density zone);

\item SFXTs on the right (highly eccentric orbit, longer quiescence
  intervals), with the NS grazing the coloured area at periastron (as
  IGR\,J00370+6122);

\item Intermediate systems if the NS is inside the narrow transition zone;

\item sgXBs: the NS is always inside the coloured area.

\end{itemize}

The observed division between sgXBs (persistent sgXBs, SFXTs, regular 
outbursters) is therefore naturally explained by simple geometrical differences
in the orbital configurations. \\

\underline{IGR\,J18483-0311: an intermediate SFXT?}

X-ray properties of this system were pointing towards an SFXT
\citep{sguera:2007}, however it exhibits an unusual behaviour: its
outbursts last for a few days (to compare to hours for classical
SFXTs), and the ratio $L_{max}/L_{min} \sim 10^3$ (its quiescence is
therefore at a higher level than the ratio $\sim 10^4$ for classical
SFXTs). Moreover, its orbital period $\Porb$=18.5d is low compared to
classical SFXTs (with large/eccentric orbits). Finally, its orbital
period and spin period ($\Pspin$=21.05s) located it inbetween Be and
sgXBs in the Corbet Diagramme, therefore in a ambiguous position.
\cite{rahoui:2008a} have identified the companion star of this system
as a B0.5Ia supergiant, unambiguously showing that this system was an
SFXT.  Furthermore, they suggest that this system could be the first
firmly identified intermediate SFXT, characterized by short, eccentric
orbit (e between 0.4 and 0.6 according to \citeauthor{rahoui:2008a}
\citeyear{rahoui:2008a}), and long outbursts... This "intermediate"
SFXT nature would explain its unusual characteristics among
"classical" SFXTs.

\subsection{Obscured HMXBs}

  \subsubsection{IGR~J16318-4848 --An extreme case--:}

  IGR~J16318-4848 was the first source discovered by IBIS/ISGRI on
  {\it INTEGRAL} on 29 January 2003 \citep{courvoisier:2003}, with a
  $2 \amin$ uncertainty.  {\it XMM-Newton} observations revealed a
  comptonised spectrum exhibiting an unusually high level of
  absorption: $\nh \sim 1.84 \times 10^{24} \cmmoinsdeux$
  \citep{matt:2003}.  The accurate localisation by {\it XMM-Newton}
  allowed \cite{filliatre:2004} to rapidly trigger ToO photometric and
  spectroscopic observations in optical/NIR, leading to the
  confirmation of the optical counterpart \citep{walter:2003} and to
  the discovery of the NIR one \citep{filliatre:2004}.  The extremely
  bright NIR source (B>25.4+/-1; I=16.05+/-0.54, J\,$= 10.33\pm 0.14$;
  H\,$=8.33\pm 0.10$ and Ks\,$=7.20 \pm 0.05 \mags$) exhibits an
  unusually strong intrinsic absorption in the optical ($A_v = 17.4
  \mags$), 100 times stronger than the interstellar absorption along
  the line of sight ($A_v = 11.4 \mags$), but still 100 times lower
  than the absorption in X-rays.  This led \cite{filliatre:2004} to
  suggest that the material absorbing in X-rays was concentrated
  around the compact object, while the material absorbing in
  optical/NIR was enshrouding the whole system.  The NIR spectroscopy
  in the $0.95-2.5 \microns$ domain allowed to identify the nature of
  the companion star, by revealing an unusual spectrum, with many
  strong emission lines:

\begin{itemize}

\item H, He\rm{I} (P-Cyg) lines, characteristic of dense/ionised wind at a
velocity of 400 km/s,

\item He\rm{II} lines: the signature of a highly excited region,

\item $[$Fe\rm{II}$]$: reminiscent of shock heated matter,

\item Fe\rm{II}: emanating from media of densities > $10^5-10^6$ cm$^{-3}$,

\item Na\rm{I}: coming from cold/dense regions.

\end{itemize}

All these lines are originating from a highly complex and stratified
circumstellar environment of various densities and temperatures,
suggesting the presence of an envelope and strong stellar outflow
responsible for the absorption. Only luminous early-type stars such as
supergiant sgB[e] show such extreme environments, and
\cite{filliatre:2004} concluded that IGR~J16318-4848 was such an unusual
HMXB, hosting a sgB[e] with characteristic luminosity of
$10^6 \Lsol$ and mass of $30 \Msol$ (see also \citeauthor{chaty:2005a} \citeyear{chaty:2005a}).

The question of this huge absorption was still pending, 
and only MIR observations would allow to solve this question,
and understand its origin.
By combining these optical and NIR data with new MIR observations, and
fitting these observations with a model of a sgB[e] companion star,
\cite{rahoui:2008} showed that IGR~J16318-4848 was exhibiting a MIR
excess (see Figure \ref{figure:igrj16318-igrj17544}, left panel), that they
interpreted as being due to the strong stellar outflow emanating from
the sgB[e] companion star.  They found that the companion star had a
temperature of $\Tstar=22200$\,K and radius $\Rstar = 20.4 \Rsol = 0.1$\,a.u., 
consistent with a supergiant star, and
an extra component of temperature T $=1100$\,K and radius R\,$= 11.9
\Rstar = 1 a.u.$, with A$_v = 17.6 \mags$. 
%
%
Recent MIR spectroscopic observations with VISIR at the VLT showed
that the source was exhibiting strong emission lines of H, He, Ne, PAH, Si,
proving that the extra absorbing component was made of dust and
cold gas.

By taking a typical orbital period of 10 days and a mass of the
companion star of $20 \Msol$, we obtain an orbital separation of $50
\Rsol$, smaller than the extension of the extra component of dust/gas
($= 240 \Rsol$),
suggesting that this component enshrouds the whole binary system, as
would do a cocoon of gas/dust (see Figure \ref{figure:obscured-sfxt},
left panel).

In summary, IGR~J16318-4848 is an HMXB system, located at a distance
between 1 and 6 kpc, hosting a compact object (probably a NS)
and a sgB[e] star (it is therefore the second HMXB with a sgB[e] star,
after CI Cam; see \citeauthor{clark:1999} \citeyear{clark:1999}). The most
striking facts are (i) the compact object seems to be surrounded by
absorbing material and (ii) the whole binary system seems to be
surrounded by a dense and absorbing circumstellar material envelope or
cocoon, made of cold gas and/or dust. This source exhibits such extreme
characteristics that it might not be fully representative of the other
obscured sources.

\subsection{The Grand Unification: different geometries, different scenarios}

There seems to be a continuous trend, from classical and/or absorbed
sgHMBs, to classical SFXTs. We outline in the following this trend.

\begin{figure*}[!ht]
\centering
\includegraphics[height=.31\textheight,angle=-90]{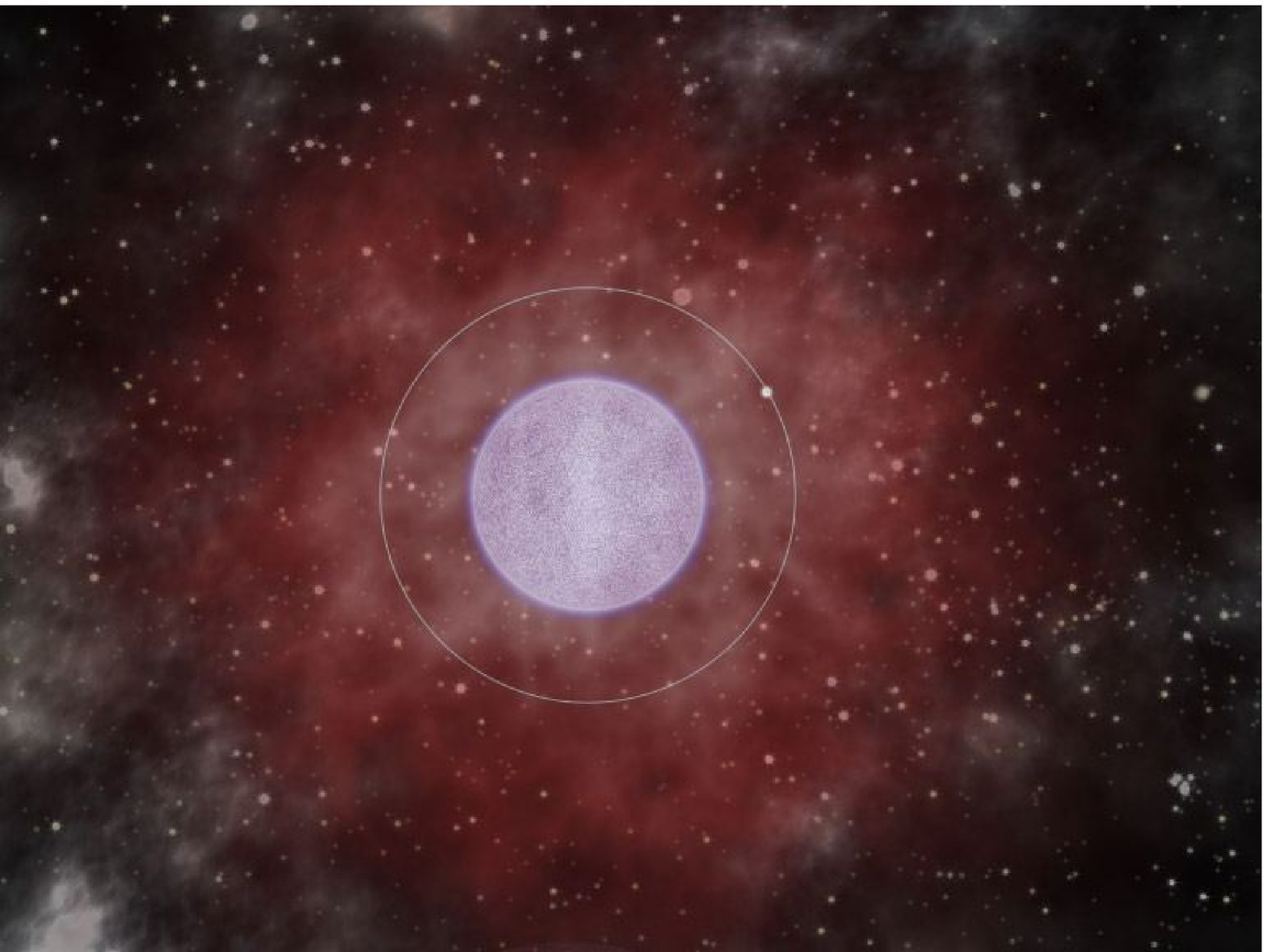}
\includegraphics[height=.31\textheight,angle=-90]{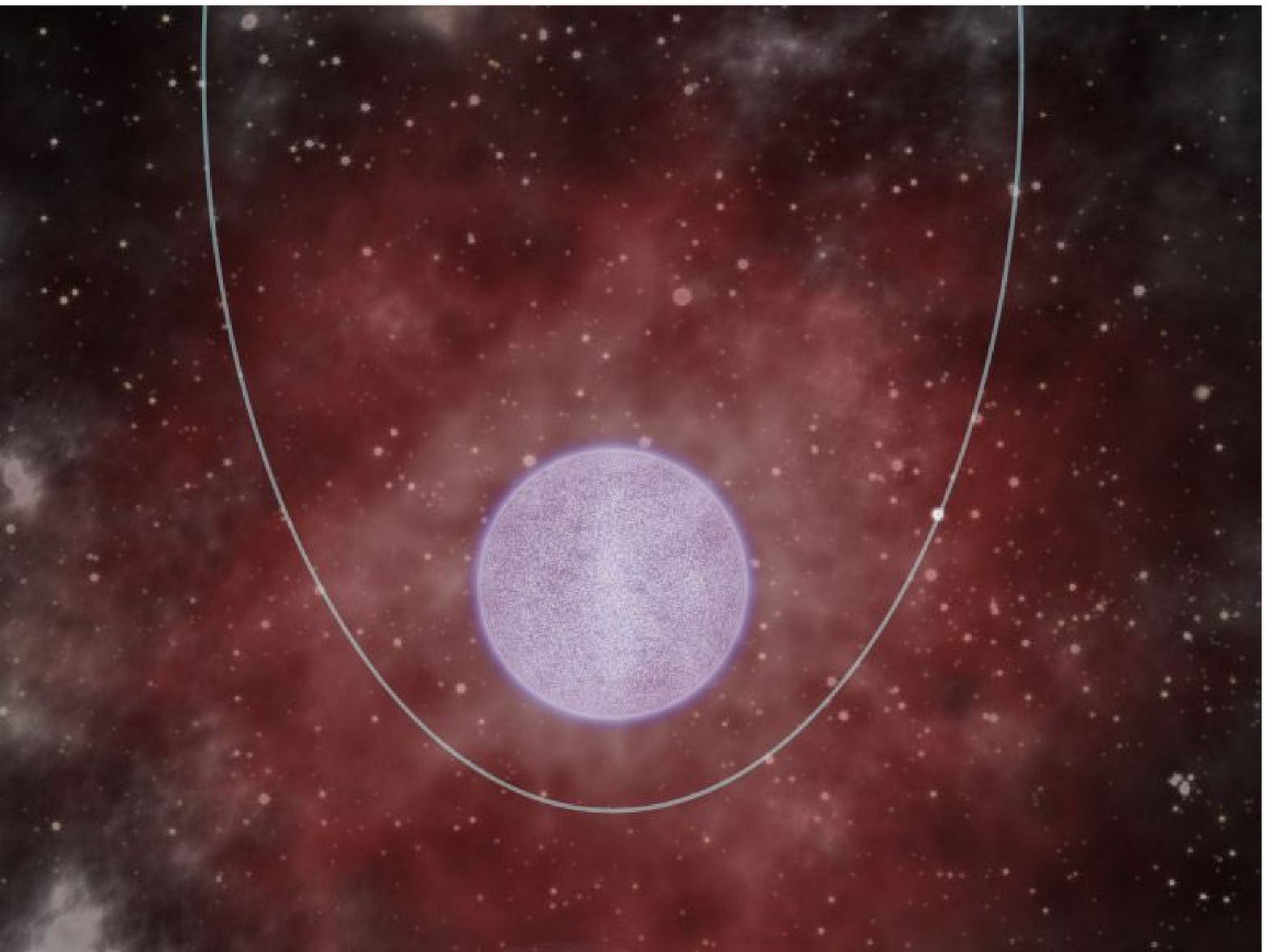}
\caption[Scenarios illustrating both 2 types of {\it INTEGRAL}
sources]{Scenarios illustrating two possible configurations of {\it
    INTEGRAL} sources: a NS orbiting a supergiant
  star on a circular orbit (left image); and on an eccentric orbit
  (right image), accreting from the clumpy stellar wind of the
  supergiant.  The accretion of matter is persistent in the case of
  the obscured sources, as in the left image, where the compact object
  orbits inside the cocoon of dust enshrouding the whole system. On
  the other hand, the accretion is intermittent in the case of SFXTs,
  which might correspond to a compact object on an eccentric orbit, as
  in the right image.  A 3D animation of these sources is available on
  the
  website:\\  
{\em http://www.aim.univ-paris7.fr/CHATY/Research/hidden.html}
}
  \label{figure:obscured-sfxt}
\end{figure*}

\begin{itemize}

\item In "classical" sgXBs, the NS is orbiting at a few
  stellar radii only from the star. The absorbed (or obscured) sgXBs (like
  IGR\,J16318-4848) are classical sgXBs hosting NS constantly
  orbiting inside a cocoon made of dust and/or cold gas, probably
  created by the companion star itself. These systems therefore exhibit a
  persistent X-ray emission.  The cocoon, with an extension of $\sim
  10 \Rstar = 1$\,a.u., is enshrouding the whole binary system. The NS
  has a small and circular orbit (see Figure
  \ref{figure:obscured-sfxt}, left panel).

\item In "Intermediate" SFXT systems (such as IGR\,J18483-0311), the
  NS orbits on a small and circular/excentric orbit, and it is only
  when the NS is close enough to the supergiant star that accretion
  takes place, and that X-ray emission arises.

\item In "classical" SFXTs (such as IGR\,J17544-2619), the NS orbits on
  a large and excentric orbit around the supergiant star, and exhibits some
  recurrent and short transient X-ray flares, while it comes close to
  the star, and accretes from clumps of matter coming from the wind of
  the supergiant.  Because it is passing through more diluted medium,
  the $\frac{Lmax}{Lmin}$ ratio is higher for "classical" SFXTs than for
  "intermediate" SFXTs (see Figure \ref{figure:obscured-sfxt}, right
  panel).

\end{itemize}

Although this scenario seems to describe quite well the characteristics
currently seen in sgXBs, we still need to identify the nature of
many more sgXBs to confirm this scenario, and in particular the
orbital period and the dependance of the column density with the phase
of the binary system.

\subsubsection{Obscured Be X-ray binary systems???}

Now that we have described the obscured sgXBs, we point out that we
might have found the most highly absorbed and distant Be star in a
HMXB system: AX J1749.1-2733.  This system is an HMXB pulsar, with a
$\Porb$=185d and a $\Pspin$=66s, and flares lasting for 12 days. Its
characteristics are therefore typical of Be star in HMXB pulsars.
\cite{zurita-heras:2008} have identified its optical counterpart: it
is a Be star exhibiting a high level of absorption ($\nh = 2 \times
10^{23} \cmmoinsdeux$), located far away in our Galaxy ($> 8.5 \kpc$),
probably similar to the BeXB 2S\,1845-024.

\subsubsection{Population synthesis models:}

These sources revealed by {\it INTEGRAL}, namely the supergiant HMXBs, will
allow to give constraints and finally to better understand the
formation and evolution of binary systems, by comparing them to
numerical study of LMXB/HMXB population synthesis models.  Many
parameters do influence the various evolutions of these systems:
differences in size, orbital period, ages, accretion type, and stellar
endpoints... Moreover, stellar and circumstellar properties also
influence the evolution of high-energy binary systems, made of two
massive components usually born in rich star forming regions.  Finally, these
new systems might represent a precursor stage of what is known as the
"Common envelope phase" in the evolution of LMXB/HMXB systems.  

These sources are also useful to look for massive stellar
"progenitors", for instance giving birth to coalescence of compact
objects, through NS/NS or NS/BH collisions. They would then become
prime candidate for gravitational wave emitters, or even to short/hard
$\gamma$-ray bursts.

\section{Conclusions and perspectives...}

The {\it INTEGRAL} satellite has tripled the total number of Galactic
sgXBs, constituted of a NS orbiting around a supergiant
star. Most of these new sources exhibit a large $\nh$ and long
$\Pspin$ (~1ks): they are slow and absorbed X-ray pulsars.  The
influence of the local absorbing matter on periodic modulations is
different for sg or BeXBs: sgOB or BeXBs are segregated in
different parts of $\nh$-$\Porb$ or $\nh$-$\Pspin$.

{\it INTEGRAL} revealed 2 new types of sources.  First, the SFXTs
(Supergiant Fast X-ray Transients) exhibiting short and strong X-ray
flares, with a peak flux of 1 Crab during 1--100s, every $\sim 100$ days.
These short and intense flares can be explained by accretion through
clumpy winds.  Second, the obscured HMXBs are composed of supergiant
stellar companions exhibiting a strong intrinsic absorption.  They are
X-ray pulsars with persistent emission, and long $\Pspin$.  The NS is
deeply embedded in the dense stellar wind, which forms a dust cocoon
enshrouding the whole binary system.

These results show the existence in our Galaxy of a dominant
population of a previously rare class of high-energy binary systems:
supergiant HMXBs, some exhibiting a high intrinsic absorption
(\citeauthor{chaty:2008} \citeyear{chaty:2008};
\citeauthor{rahoui:2008} \citeyear{rahoui:2008}).  A careful study of
this population, recently revealed by {\it INTEGRAL}, will provide a
better understanding of the formation and evolution of short-living
HMXBs.  Furthermore, stellar population models will henceforth have to
take these objects into account, to assess a realistic number of
high-energy binary systems in our Galaxy. Our final word is that only
a multiwavelength study reveal the nature of these obscured
high-energy sources.

The recently successfully launched GLAST satellite (Gamma-ray Large
Area Space Telescope, operating in the 10 keV - 300 GeV energy range)
will certainly discover such new and unexpected objects.  Indeed,
although these sgXB sources do not seem to be GeV Emitters, there are
peculiar members of this family which emit at these energy ranges, for
instance LSI +61 303 (a "disguised" pulsar, \citeauthor{dubus:2006a}
\citeyear{dubus:2006a}), LS 5039 
(\citeauthor{paredes:2000} \citeyear{paredes:2000};
\citeauthor{aharonian:2005} \citeyear{aharonian:2005}), 
and Cygnus X-1\footnote{the only sgXB hosting
  a confirmed BH.}  \citep{albert:2007}.  This high energy satellite
has a huge potential: 200 individual sources have been detected by
EGRET... we expect that thousands will be detected by GLAST!

\acknowledgments SC would like to thank the organisers of this
successful and exciting school for their invitation to give this
lecture on the not less exciting newly discovered {\it INTEGRAL}
sources in a nice place, fruitful for scientific discussions between
students and professors. Gracias Paula y Gustavo! \\
Y como lo decia Alberto Ludwig Urquieta en 1926,
"El universo es tremendamente creativo, lo que nos obliga a abrirnos a
lo desconocido...".




\end{document}